\documentclass[10pt,twocolumn,letterpaper]{article}

\usepackage{cvpr}
\usepackage{times}
\usepackage{epsfig}
\usepackage{graphicx}
\usepackage{amsmath}
\usepackage{amssymb}
\usepackage{booktabs}  
\usepackage{subcaption}  
\usepackage{multirow}
\usepackage[pagebackref=true,breaklinks=true,letterpaper=true,colorlinks,bookmarks=false]{hyperref}

\cvprfinalcopy 
\def\cvprPaperID{****} 

\ifcvprfinal\fi
\begin{document}

\title{MMAudioReverbs: Video‑Guided Acoustic Modeling for Dereverberation\\and Room Impulse Response Estimation}  
\author{
Akira Takahashi$^{1}$ \quad
Ryosuke Sawata$^{2}$ \quad
Shusuke Takahashi$^{1}$ \quad
Yuki Mitsufuji$^{1,2}$\\
$^{1}$Sony Group Corporation \quad
$^{2}$Sony AI}

\maketitle
\thispagestyle{empty}

\begin{abstract}
Although recent video-to-audio (V2A) models excelled at synthesizing semantically plausible sounds from visual inputs, they do not explicitly model room-acoustic effects such as reverberation or room impulse responses (RIRs), and thus offer limited controllability over these effects. 
However, we hypothesize that such V2A models implicitly have semantic knowledge of the relationship between spatial audio and the corresponding vision cues.
In this paper, we revisit a V2A model for the sake of the above, and propose the way to utilize the pretrained model as prior for physically grounded room-acoustic processing.
Based on one of the state-of-the-art V2A models, MMAudio, we propose MMAudioReverbs that is a unified framework dealing with i) dereverberation and ii) room impulse response (RIR) estimation without network architectural modification, and fine-tuned on a small dataset.
Experimental results showed that audio and visual cues respectively have advantage depending on the type of physical room acoustics.
It implies that foundation V2A models can be used for physically grounded room-acoustic analysis.
\end{abstract}

\section{Introduction} \label{sec:intro}

Modeling physical room acoustics is essential for numerous audio and multimedia applications, including speech enhancement, virtual acoustics, and realistic audio generation for videos. 
Existing video-to-audio (V2A) methods~\cite{cheng2025taming,chen2024multifoley} can synthesize plausible sounds; however, they do not explicitly model room-acoustic effects such as reverberation or room impulse responses (RIRs), and thus offer limited controllability over these effects. 

Some researchers have explored multimodal approaches~\cite{Ratnarajah_2024_CVPR,majumder2022fewshot,luo2022learning}, including video-conditioned methods, but these approaches are limited in their ability to exploit visual cues that encode physical scene properties such as room geometry, spatial layout, and material characteristics.
To address this limitation, we propose a way to leverage the potential of existing V2A models for physically grounded acoustic processing.
Inspired by MMAudioSep~\cite{takahashi2025mmaudiosep}, we hypothesize that state-of-the-art V2A foundation models may implicitly encode relationships between visual cues and room-acoustic properties.
Therefore, we adopt one of the state-of-the-art V2A models, MMAudio~\cite{cheng2025taming}, as our backbone, and enable it to deal with two core tasks related to physical room acoustics, i.e., i) dereverberation and ii) RIR estimation, without architectural modification.
Specifically, just by fine-tuning MMAudio upon the small dataset, our new model, named \emph{MMAudioReverbs}, can deal with dereverberation and RIR estimation.
Note that no modifications of network architecture are necessary.
This is motivated by the hypothesis that the backbone model, MMAudio, may implicitly encode information about scene layout, object arrangement, and source–receiver relationships. Therefore, we leverage pretrained MMAudio as a source of physical priors for physically grounded acoustic estimation.

\begin{figure}[t]
  \centering
  \begin{minipage}[b]{0.49\linewidth}
    \centering
    \includegraphics[width=0.95\linewidth]{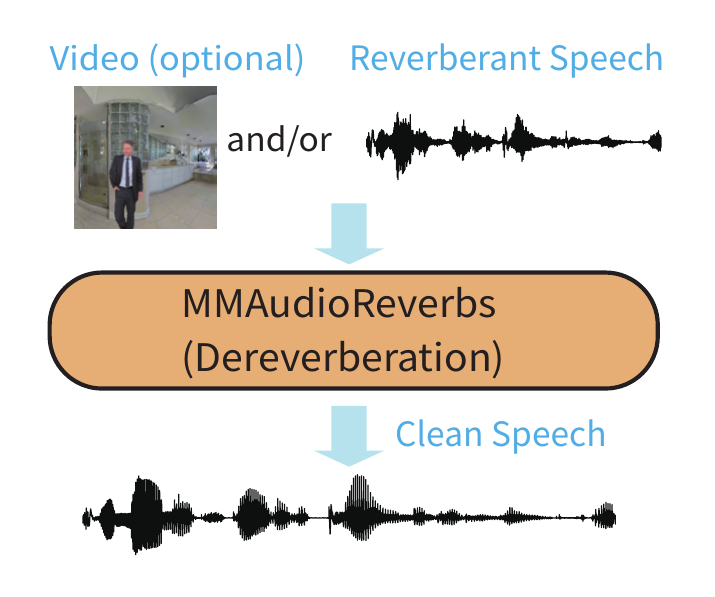}
    \vspace{-1mm}
    \\(a) Dereverberation
  \end{minipage}
  \hfill
  \begin{minipage}[b]{0.49\linewidth}
    \centering
    \includegraphics[width=0.95\linewidth]{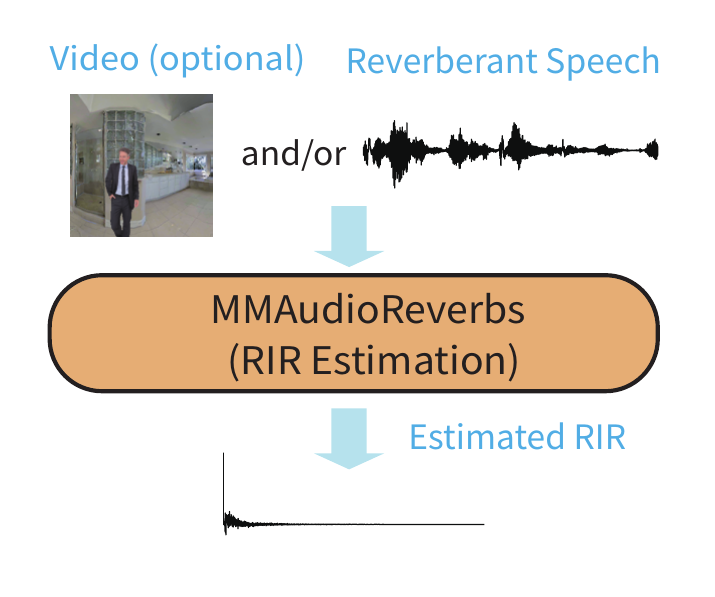}
    \vspace{-1mm}
    \\(b) RIR Estimation
  \end{minipage}

  \vspace{2mm}
  \caption{
    Outline of MMAudioReverbs that can handle two core tasks related to physical room acoustics, dereverberation and RIR estimation. 
  }
  \label{fig:concept}
  \vspace{-4mm}
\end{figure}

Figure~\ref{fig:concept} provides a conceptual overview of this idea, illustrating how MMAudioReverbs is conditioned on visual inputs to support both dereverberation and RIR estimation.

Experimental results show that visual information contributes to more stable dereverberation behavior and improves the physical interpretability of estimated RIRs, particularly in terms of early energy characteristics that are visually consistent with the underlying scene. 

In summary, we demonstrate that a pretrained multimodal V2A foundation model can be directly repurposed, without architectural modification, for physically grounded room-acoustic tasks such as dereverberation and RIR estimation, and that visual information functions as a physical prior that complements acoustic evidence in these settings.

\begin{figure*}[t]
  \centering

  \begin{minipage}[t]{0.50\textwidth}
    \vspace{0pt}
    \centering
    \includegraphics[width=0.99\linewidth]{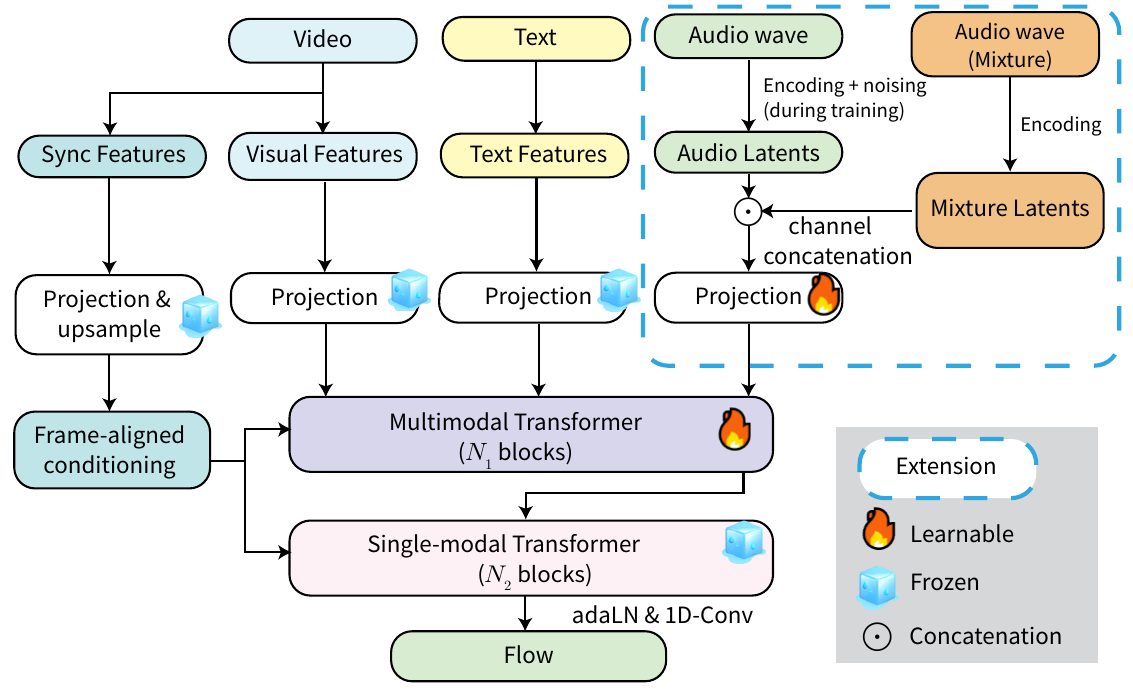}
    \smallskip
    \\(a) MMAudio Architecture
  \end{minipage}
  \hfill
  \begin{minipage}[t]{0.24\textwidth}
    \vspace{0pt}
    \centering
    \includegraphics[width=0.99\linewidth]{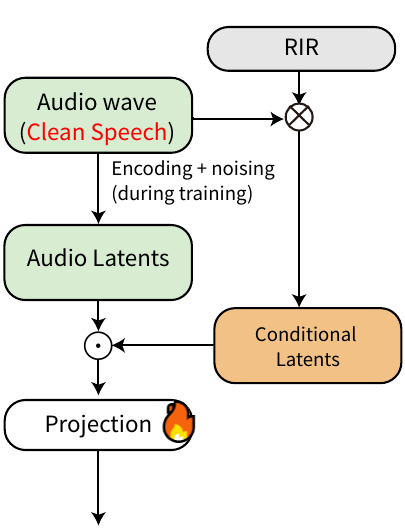}
    \smallskip
    \vspace{0mm}
    (b) Dereverberation
  \end{minipage}
  \hfill
  \begin{minipage}[t]{0.24\textwidth}
    \vspace{0pt}
    \centering
    \includegraphics[width=0.99\linewidth]{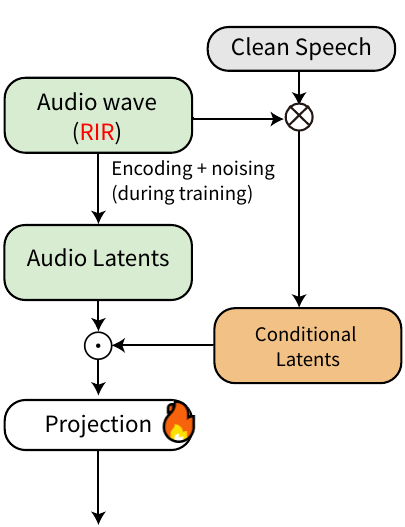}
    \smallskip
    \vspace{0mm}
    (c) RIR Estimation
  \end{minipage}
  \vspace{2mm}
  \caption{
    MMAudio backbone and its task-specific reinterpretation for dereverberation and RIR estimation.
(a) MMAudio architecture (16\,kHz).
    By replacing (b) Dereverberation or (c) RIR estimation with the part highlighted by light blue dashed in (a), MMAudioReverbs adaptively deal with these two tasks.
  }
  \label{fig:architecture_and_conditioning}
\end{figure*}
\section{Related Work}

\subsection{V2A Generation and Multimodal Foundations}
Recent progress in audio-visual learning has led to powerful V2A generation models that synthesize realistic sounds conditioned on visual inputs~\cite{chen2024multifoley,cheng2025taming}. 
However, these models primarily target perceptual and content realism, rather than explicit modeling of physical room acoustics such as reverberation or RIR. 

\subsection{Room Acoustics with Audio and Visual Cues}
Speech dereverberation and RIR estimation have been extensively studied in the context of audio-only signal processing and learning-based methods, often relying on task-specific architectures tailored to particular acoustic objectives. 
More recent work has begun to incorporate visual information in room-acoustic modeling, including geometry-aware RIR estimation, visually guided sound propagation, and material-aware acoustic modeling~\cite{luo2022learning,Saad_2025_ICCV,Singh_2021_ICCV}.

In contrast to these approaches, which introduce task-specific architectures to encode geometric priors, we explore a complementary perspective by evaluating whether visual priors can be implicitly leveraged by a pretrained multimodal V2A foundation model without task-specific architectural specialization.

\section{Method}

\subsection{Backbone and Conditioning Framework}
\label{sec:Backbone}
Our approach builds on MMAudio~\cite{cheng2025taming}, a pretrained V2A foundation model for video-to-audio generation, and fine-tunes its multimodal backbone for video-guided room-acoustic tasks.
As illustrated in Fig.~\ref{fig:architecture_and_conditioning}, the model provides a unified multimodal conditioning interface. 
Within this framework, conditioning signals modulate latent trajectories to produce acoustically plausible outputs consistent with the observed scene.

\subsection{Unified Flow Formulation for Acoustic Tasks}\label{sec:flow}
MMAudio operates in the latent space of a pretrained audio VAE (Variational Autoencoder) and is trained using a flow-matching objective. 
Importantly, the same flow dynamics are reused across tasks without reparameterization, enabling both inverse mappings and conditional generation within a single unified formulation.

Therefore, we instantiate different room-acoustic tasks by reinterpreting the roles of conditioning signals and target latent trajectories within the same flow formulation, while sharing the same learnable parameters and architectural components across tasks. 
Dereverberation models the conditional mapping from reverberant to clean speech, suppressing acoustically inconsistent reflections. 
RIR estimation conditionally generates room impulse response consistent with the input reverberant audio.

In summary, once MMAudioReverbs is built, it can deal with two core physical acoustic tasks (i.e., dereverberation and RIR estimation), just by replacing the upper right part in Fig.~\ref{fig:architecture_and_conditioning} (a) with the part in Fig.~\ref{fig:architecture_and_conditioning} (b) or Fig.~\ref{fig:architecture_and_conditioning} (c).

\section{Experiments}
\label{sec:experiments}
To study the effects of pretrained weights, we trained two models: (i) from scratch and (ii) fine-tuned from pretrained MMAudio. At inference time, we evaluated each model under two conditioning settings: audio-only and audio+vision.
Note that sampling rate used in our experiments was set at 16 kHz since the SoundSpaces-Speech dataset~\cite{chen22soundspaces2,chen22dereverb} is provided at 16 kHz.

\subsection{Setup}
\paragraph{Pretrained Model Setup.}
We utilized pretrained MMAudio to initialize our model and then fine-tuned it for our MMAudioReverbs.
Note that we also report the results of MMAudioReverbs trained from scratch to confirm the effectiveness of the fine-tuning.

\vspace{-5mm}
\paragraph{Datasets.}
We used the SoundSpaces‑Speech dataset, where panoramic RGB images are cropped into square views with a 120° field of view and used as visual conditioning inputs.
\vspace{-5mm}
\paragraph{Fine-tuning Setup.}
We trained our models using 2.56 sec segments and 20k training steps.
\vspace{-5mm}
\paragraph{Vocoder Configuration.}
Waveforms were reconstructed from the predicted latent representations using BigVGAN~\cite{lee2023bigvgan}. Due to task‑dependent signal characteristics, the vocoder was trained separately for each task: clean speech from LibriSpeech\footnote{https://www.openslr.org/12/} for dereverberation and RIRs from the SoundSpaces‑Speech training set for RIR estimation.
\vspace{-5mm}
\paragraph{Evaluation Setup.}
For RIR estimation, both input and predicted RIRs were evaluated using fixed 2.56 sec windows, with input speech obtained by extracting the central segment of each test utterance, whereas dereverberation was evaluated on variable‑length audio at its original duration. 
Classifier‑free guidance (CFG) was disabled during inference. We found that it introduces additional generative variability, degrading \emph{estimation} performance. 

\subsection{Results}

\paragraph{Metrics.}
For dereverberation, we report Reverberation Time (RT60), Reverberation Time Error (RTE), Speech-to-Reverberation Modulation Energy Ratio (SRMR), and Deep Noise Suppression Mean Opinion Score (DNSMOS).
RTE is the absolute error between the estimated RT60 values of the output and reference signals, computed using a blind reverberation time estimator~\cite{chen22vam}. 
DNSMOS predicts perceptual mean opinion scores (MOS).
For RIR estimation, we report parameter‑level errors in RT60, Early Decay Time (EDT), and Direct-to-Reverberant Ratio (DRR), computed as absolute error between parameters estimated from the predicted and reference RIRs.

\vspace{-5mm}
\paragraph{Dereverberation.}
Table~\ref{tb:exp_dereverb} shows that our method substantially improved dereverberation quality over other comparative methods. 
In particular, MMAudioReverbs fine-tuned from the pretrained initialization reduces RTE compared to the corresponding MMAudioReverbs trained from scratch.
This implies that pretrained multimodal representations provide a beneficial initialization for dereverberation. 
Note that our outputs can exceed the ‘Clean’ reference on some metrics, since the ‘Clean’ signals in the dataset may still contain background noise and residual reverberation, whereas our model can further suppress these components.
However, MMAudioReverbs achieved similar dereverberation performance with audio-only and audio+vision conditioning when initialized identically.
This suggests that audio features alone may be sufficient for dereverberation, since both settings include acoustic conditioning.

\begin{figure}[t]
  \centering

    \centering
    \includegraphics[width=1.0\linewidth]{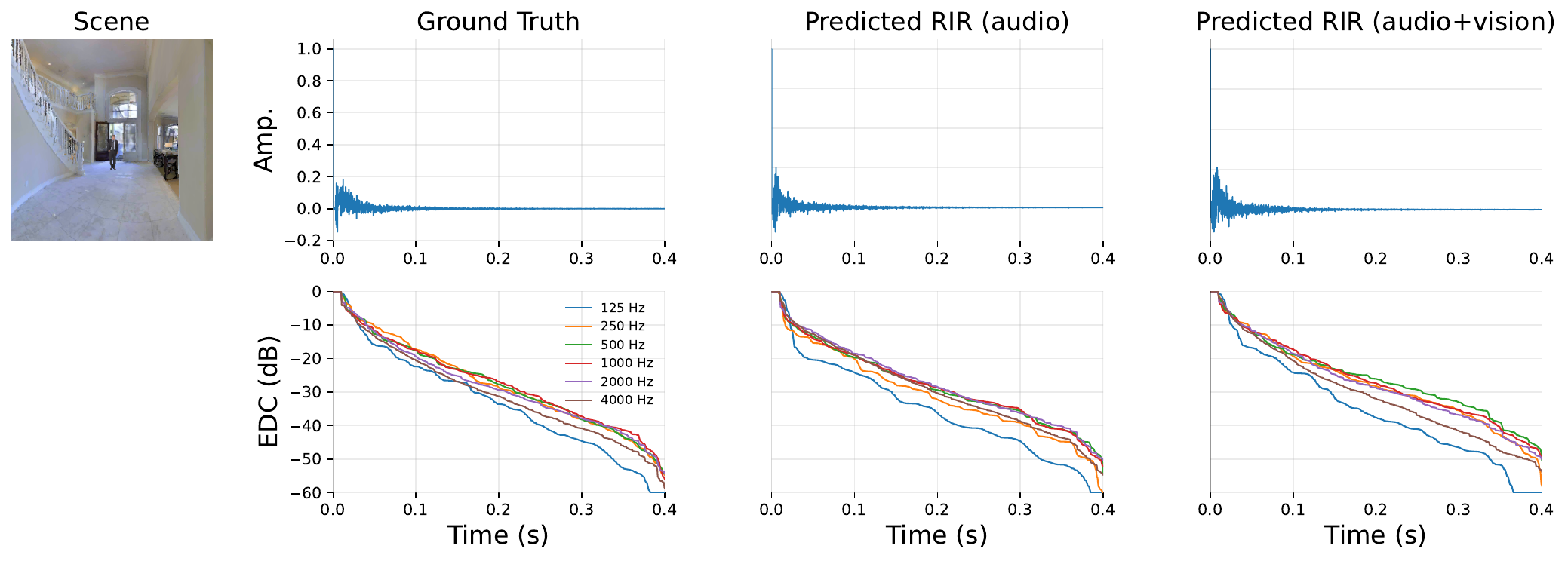}

  \caption{Visualization examples of RIR estimation. Top: waveforms of the Ground Truth and predicted RIRs under different conditioning settings (audio, audio+vision). Bottom: corresponding energy decay curves (EDC) computed from each RIR.}
  \label{fig:viz_rir}
\end{figure}

\vspace{-5mm}
\paragraph{RIR Estimation.}
As shown in Table~\ref{tb:exp_rir}, audio‑only conditioning yielded lower errors for late‑reverberation metrics such as RT60, reflecting the strong dependence of decay characteristics on temporal acoustic evidence. 
In contrast, incorporating visual information leads to reduced DRR errors in several settings, indicating that visual cues can provide complementary information for modeling early energy and direct‑path dominance.
We also observed that fine-tuning from the pretrained model improves RIR estimation accuracy compared to training from scratch, yielding lower errors across most metrics. 
This suggests that pretrained multimodal representations provide a strong starting point for room-acoustic modeling.
Figure \ref{fig:viz_rir} visualizes representative RIR estimation examples under different conditioning settings.

\begin{table}[t]
\setlength\tabcolsep{3pt}
\caption{
    Experimental results.
    Higher is better for $\uparrow$, and lower is better for $\downarrow$.
    `A' and `A+V' denote the modalities, i.e., (audio) and (audio+vision), used for the corresponding method.
}
 \label{tb:exp_results}
 \begin{subtable}[t]{\columnwidth}
\subcaption{
  Dereverberation: VIDA~\cite{chen22dereverb} results are reproduced using the official repository.
}
 \label{tb:exp_dereverb}
 \centering
 \resizebox{\columnwidth}{!}{
\begin{tabular}{l c ccc ccc}
\toprule
 & \multirow[c]{2}{*}{\textbf{Modality}} & \multirow[c]{2}{*}{\textbf{SRMR}$\uparrow$}
 & \multirow[c]{2}{*}{\textbf{RT60} (ms)$\downarrow$}
 & \multirow[c]{2}{*}{\textbf{RTE} (ms)$\downarrow$}
 & \multicolumn{3}{c}{\textbf{DNSMOS}$\uparrow$} \\
\cmidrule(lr){6-8}
  &&&&& \textbf{SIG} & \textbf{BAK} & \textbf{OVRL} \\
\midrule
Clean & -- & 7.26 & 39.4  & --    & 3.55 & 3.87 & 3.19 \\
Reverberant & -- & 4.75 & 403.1 & 363.9 & 2.53 & 2.71 & 2.09 \\
WPE~\cite{nakatani2010speech} & A & 5.97 & 137.2 & 127.3 & 2.78 & 3.07 & 2.34 \\
VIDA~\cite{chen22dereverb} & A+V & 6.54 & 78.2  & 56.2  & 3.05 & 3.55 & 2.62 \\
\midrule
\multirow[c]{2}{*}{Ours (Scratch)} & A & 7.22 & 30.1 & 29.4 & \textbf{3.57} & 3.94 & \textbf{3.24} \\
\phantom{Ours (Scratch)} & A+V & 7.24 & 30.3 & 29.7 & \textbf{3.57} & 3.94 & 3.23 \\
\multirow[c]{2}{*}{Ours (Finetune)} & A & 7.27 & \textbf{27.1} & \textbf{28.7} & \textbf{3.57} & 3.95 & \textbf{3.24} \\
\phantom{Ours (Finetune)} & A+V & \textbf{7.29} & 27.2 & 28.9 & \textbf{3.57} & \textbf{3.96} & \textbf{3.24} \\
\bottomrule
\end{tabular}
}
\end{subtable}

\vspace{1em}

\begin{subtable}[t]{\columnwidth}
\caption{ 
    RIR Estimation: $\Delta$ denotes absolute error. All baseline results are quoted from AV-RIR~\cite{Ratnarajah_2024_CVPR} under their evaluation setting.
}
 \label{tb:exp_rir}
 \centering
 \resizebox{\columnwidth}{!}{
 \begin{tabular}{l c ccc}
\toprule
 & \textbf{Modality} & \textbf{$\Delta$RT60} (ms)$\downarrow$ & \textbf{$\Delta$DRR} (dB)$\downarrow$
& \textbf{$\Delta$EDT} (ms) $\downarrow$  \\
\midrule
Image2Reverb \cite{Singh_2021_ICCV}   & V & 131.7 & 4.94 & 382.1 \\
FiNS \cite{steinmetz2021fins}    & A & 87.7 & 3.30 & 235.7 \\
S2IR-GAN \cite{10094770}   & A & 63.1 & 3.04 & 168.3 \\
\multirow[c]{2}{*}{AV-RIR \cite{Ratnarajah_2024_CVPR}} & A & 88.8 & 2.96 & 122.4 \\
   & A+V & 40.2 & 1.76 & 77.2 \\
\midrule
\multirow[c]{2}{*}{Ours (Scratch)} & A & 78.9 & 2.89 & 81.7 \\
 & A+V & 100.8 & 2.74 & 56.2 \\
\multirow[c]{2}{*}{Ours (Finetune)}   & A & 51.6 & 2.40 & 41.9 \\
 & A+V & 60.0 & 2.36 & 47.5 \\
\bottomrule
\end{tabular}
}
\end{subtable}
\end{table}

\subsection{Discussion}

Visual conditioning provides limited improvement for acoustic characteristics dominated by late reverberation such as decay-related metrics (see the difference between ``A'' and ``A+V'' of ours in Table~\ref{tb:exp_dereverb}), while offering clearer benefits for early energy characteristics (see the difference between ``A'' and ``A+V'' of ours in Table~\ref{tb:exp_rir}). 
This behavior highlights a key insight for multimodal acoustic modeling: visual cues primarily act as structural priors for early sound propagation, while late reverberation remains fundamentally constrained by acoustic evidence.
In other words, late reverberation is largely determined by temporal acoustic evidence accumulated over long durations, whereas early energy and direct-path dominance are more closely related to scene layout and source–receiver geometry, which may be partially inferred from visual cues.

In terms of the relationship between the sound source and the receiver, the source was not always visible in the input image, and thus it may make estimating source–receiver relationships from vision alone difficult.
The distance between the source and the receiver may affect the factor related to late reverberation, and thus the aforementioned ambiguity by visual cue did not improve the metrics related to late reverberation.
On the other hand, in terms of early energy-related factors such as reflection and direct sound from the source, understanding at least there is no sound source in the receiver's view angle is important.
Hence, the improvements observed in early energy-related metrics can be interpreted as evidence that implicit visual cues provide physically meaningful, though incomplete, priors for room-acoustic modeling.

\section{Conclusion}

We introduced MMAudioReverbs, a unified formulation that repurposes a pretrained multimodal V2A backbone for room-acoustic processing without architectural modification.
By utilizing the prior from the pretrained V2A foundation model, MMAudioReverbs can deal with two core physical acoustic tasks, i.e., dereverberation and RIR estimation.
Our results show that visual cues primarily act as structural priors for early sound propagation, while late reverberation remains largely constrained by acoustic evidence.

However, MMAudioReverbs does not incorporate explicit physical attributes such as scene geometry, depth, or material properties, but solely relies on RGB images and representations learned through multimodal pretraining.
In addition, the datasets used in our experiments did not consistently provide explicit visual cues or annotations for sound source locations, limiting precise modeling of source–receiver relationships.
Future work includes integrating explicit physical attributes via lightweight adapters and constructing datasets with richer source–receiver annotations to further evaluate physically grounded acoustic modeling.

{
\footnotesize
\bibliographystyle{ieee}
\bibliography{example_bib}
}

\end{document}